\documentclass[12pt]{article}
\usepackage{epsfig}

\begin{document}

\begin{titlepage}
\begin{center}\Large
{\bf \boldmath Study of the radiative decay $\phi \rightarrow
\eta \gamma$ with CMD-2 detector} \\
\end{center}

\begin{center}
R.R.Akhmetshin, E.V.Anashkin, V.S.Banzarov, L.M.Barkov, 
N.S.Bashtovoy, A.E.Bondar, D.V.Bondarev, D.V.Chernyak, 
S.I.Eidelman, G.V.Fedotovich, N.I.Gabyshev, A.A.Grebeniuk, 
D.N.Grigoriev, P.M.Ivanov, V.F.Kazanin, B.I.Khazin, I.A.Koop, 
P.P.Krokovny, L.M.Kurdadze, A.S.Kuzmin, I.B.Logashenko, 
P.A.Lukin, A.P.Lysenko, K.Yu.Mikhailov, V.S.Okhapkin, 
T.A.Purlatz, N.I.Root, A.A.Ruban, N.M.Ryskulov, A.G. Shamov, 
Yu.M.Shatunov, B.A.Shwartz, V.A.Sidorov, A.N.Skrinsky, 
V.P.Smakhtin, I.G.Snopkov, E.P.Solodov, P.Yu.Stepanov, 
A.I.Sukhanov, Yu.V.Yudin, S.G.Zverev
\\
  Budker Institute of Nuclear Physics, Novosibirsk, 630090, Russia
\end{center}
\begin{center}
           B.L.Roberts
\\
                   Boston University, Boston, MA 02215, USA
\end{center}
\begin{center}
                J.A.Thompson
\\
                University of Pittsburgh, Pittsburgh, PA 15260, USA
\end{center}
\begin{center}
                V.W.Hughes
\\
                    Yale University, New Haven, CT 06511, USA
\end{center}

\vspace{1cm}
\begin{abstract}
Using the  $1.9 \ pb^{-1}$ of data collected with the CMD-2
detector at VEPP-2M the decay mode $\phi \rightarrow \eta \gamma$,
$\eta \to \pi^+\pi^-\pi^0$ has been studied. The obtained branching ratio is 
B($\phi \rightarrow \eta \gamma) ~=~ (1.18 \pm 0.03 \pm 0.06) \%$.
\end{abstract}
\end{titlepage}

\section{Introduction}

\hspace*{\parindent}Radiative transitions between vector and
pseudoscalar mesons are very interesting for tests of the quark model, 
SU(3) symmetry and Vector Dominance Model \cite {don}. 
Recently there were extensive discussions concerning mechanisms of
SU(3) breaking, the role of anomalies and a possible admixture of glue
in mesons in radiative decays \cite{ben1,bram1,hashimoto,ben2,ball,bram2,ben3}. 
Despite numerous experimental efforts, many open questions
still exist requiring more precise measurements of the decay
probabilities. 

The radiative magnetic dipole transition of $\phi$ into
$\eta$ has been previously studied in many experiments \cite{pdg} based
on the neutral final states arising when $\eta$ decays into $\gamma\gamma$ 
or $3\pi^0$. In this work, using the general purpose CMD-2 detector
at the high luminosity Novosibirsk collider VEPP-2M, 
the $\phi \to \eta \gamma$ decay rate into the charged-neutral final 
state was measured: 
$e^+e^- \to \phi \to \eta\gamma\to \pi^+\pi^-\pi^0 \gamma$.
The study is based on a data sample  
of about $1.9 \ pb^{-1}$ corresponding to 3.5 million $\phi$ meson
decays collected in 1996 with CMD-2.

\section{Experiment}

\hspace*{\parindent}
The CMD-2 detector is described in detail elsewhere
\cite{CMD285,cmd2gen,cmdpre93}. It  is an axial field detector consisting 
of a drift chamber with about 250 $\mu$ resolution transverse to the
beam and proportional Z-chamber used for trigger, both inside a thin (0.4
$X_0$) superconducting solenoid with a field of 1 T. 

The barrel calorimeter placed outside of the solenoid consists of 892 CsI
crystals of $6\times 6\times 15$ cm$^3$ size and covers polar angles from
$46^\circ$ to $132^\circ$. The energy resolution for photons in the
CsI calorimeter is about 9\% in the energy range from 50 to 600 MeV. 

The BGO ($\mathrm{Bi_4}\mathrm{Ge_3}\mathrm{O_{12}}$) end-cap calorimeter
consists of two identical end-caps and covers forward-backward angles
from 16$^\circ$ to 49$^\circ$ and from 131$^\circ$ to 164$^\circ$. 
It is placed inside the main solenoid in the gap between the drift chamber
and the flux return yoke and operates in a 1~T longitudinal magnetic
field. Each end-cap is a compact matrix of rectangular BGO crystals of
$2.5\times2.5\times 15$ cm$^3$ size. The thickness of the calorimeter
for normally incident particles is equal to 13.4 $X_0$. The calorimeter
consists of 680 BGO crystals with a total weight of about 450~kg. The
light from each crystal is read out by vacuum phototriodes which can
operate in magnetic fields up to 2~T. The energy and angular resolution
of the end-cap calorimeter is equal to $\sigma_E/E = 4.6\%/\sqrt{E(GeV)}$
and $\sigma_{\varphi,\theta} = 2\cdot10^{-2}/\sqrt{E(GeV)}$ radians
respectively.

Data was taken in the $\phi$-meson energy range in
April -- July, 1996. It was the first run after installing the BGO end-cap 
calorimeter. The information about the data sample included in the analysis 
is presented in Table \ref{tab:rez}. The luminosity was determined using the
number of detected $e^+e^- \to e^+e^-$ events \cite{prep99}. 

\section{Selection criteria}

\hspace*{\parindent}Events with two tracks and more than one photon 
were selected using the following criteria:
\begin{itemize}
\item{ One vertex is found in the event}
\item{ Two tracks with opposite charges are reconstructed from
this vertex and there are no other tracks}
\item{ The angles of both tracks with respect to the beam 
are limited by $40^\circ< \theta <140^\circ$ to match the optimal drift
chamber coverage}
\item{ The number of photons detected in the CsI and BGO calorimeters
is more than one and less than six. The cluster in the calorimeter was
accepted as a photon when it did not match any charged track and its
energy was more than 30 MeV in the CsI calorimeter (more than 40 MeV in
the BGO calorimeter).}
\item{ The distance from each track to the beam $R_{min}~<~0.2~cm$}
\item{ The distance from the vertex to the interaction point
along the beam direction $|Z_{vert}|~<~10~cm$}
\item{ The space angle between the tracks $~\Delta \psi~<~
143^\circ$ to suppress events of $\phi\to K_SK_L$, $K_S\to\pi^+\pi^-$}
\item{ The angle between the tracks in the r-$\varphi$ plane
$~\Delta \varphi~<~172^\circ$}
\item{ The total energy of the charged particles (assuming
that both are charged pions) $\varepsilon_{\pi^+\pi^-}<520~MeV$}
\item{ The energy deposition of the photon accepted as a recoil one
is more than 250 MeV}
\end{itemize}

The cut for the total energy $\varepsilon_{\pi^+\pi^-}<520~MeV$ is to
reject events from the decay $\phi \rightarrow \pi^+\pi^-\pi^0$. From 
simple kinematical relations the total energy of charged pions from the
$\phi\rightarrow \pi^+\pi^-\pi^0$ is greater than 539 MeV 
while for $\phi \rightarrow \eta\gamma$ it is less than 522 MeV.

\section{\boldmath Kinematics of $\phi\rightarrow\eta\gamma$}
\label{sec:kin}

\hspace*{\parindent}
In this study the final state contains two charged particles and 
three photons. 

Since $\phi \rightarrow \eta \gamma$ is a two-body decay and $\eta$ is a
narrow state, the energy of the recoil photon is given by
\begin{equation}
\omega_{r} = \frac{(2E_{b})^2 - m^2_\eta}{4E_{b}},
\label{eq:wr}
\end{equation}
where $E_b$ is the beam energy, $m_\eta$ is the $\eta$ mass.
The invariant mass of two other final photons is equal to $m_{\pi^0}$.
At the $\phi$ meson peak the energy of the recoil photon equals 363 MeV.

In this analysis detection of all final particles is not necessary.
Instead, one requires the detection of both charged pions and at least
one photon, the recoil one. Using the measured momenta of charged pions,
angles $\varphi$ and $\theta$ of the recoil photon and the assumption
that the energy of the recoil photon is given by (\ref{eq:wr}),
one can reconstruct the invariant mass of all other photons in the
system, $M_{inv}$. In case of the process 
$e^+e^- \to \phi \to \eta\gamma, \eta \to \pi^+\pi^-\pi^0$
this parameter $M_{inv}=m_{\pi^0}$.

\begin{figure}
\begin{center}
\includegraphics[width=0.9\textwidth]{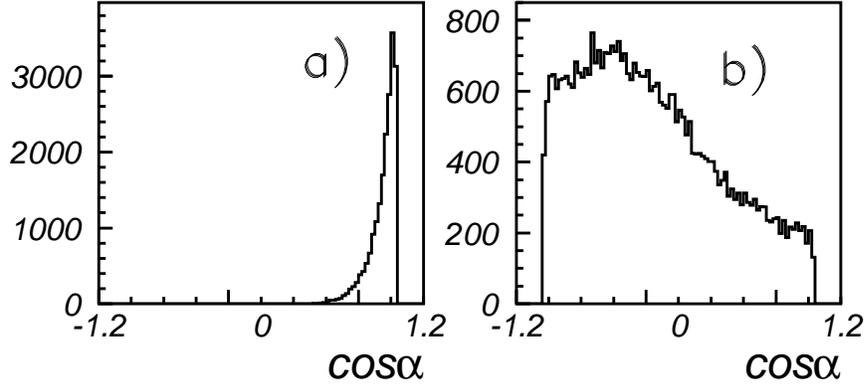}
\vspace{-6.5cm}
\caption{Distributions over cosine of the photon angle with
respect to  $-\vec{p}_{\pi^+\pi^-}$ for the simulation of
$\phi \to \eta \gamma \rightarrow \pi^+ \pi^- \pi^0 \gamma$:
{\bf a)} for the recoil photon;
{\bf b)} for one of the photons from the $\pi^0$ decay.}
\label{fig:simang}
\end{center}
\end{figure}

The recoil photon has the largest energy among three photons in the final 
state. So, the simplest way would be to take a photon
with the largest energy deposition in the ca\-lo\-rimeter. In this study,
however, we prefer not to use the energy deposition at all, 
relying instead on the position of the cluster (angles of photons). 
The Monte Carlo simulation of the decay mode studied in this work shows
that the recoil photon prefers to fly opposite to the  
total momentum of charged pions $\vec{p}_{\pi^+\pi^-}$
in contrast to the photons from the $\pi^0$ decay (Fig.~\ref{fig:simang}).
Therefore the photon whose direction is closest to  $~-\vec{p}_{\pi^+\pi^-}$
was selected as the recoil one.

\section{Data Analysis}

\begin{figure}
\begin{center}
\includegraphics[width=0.9\textwidth]{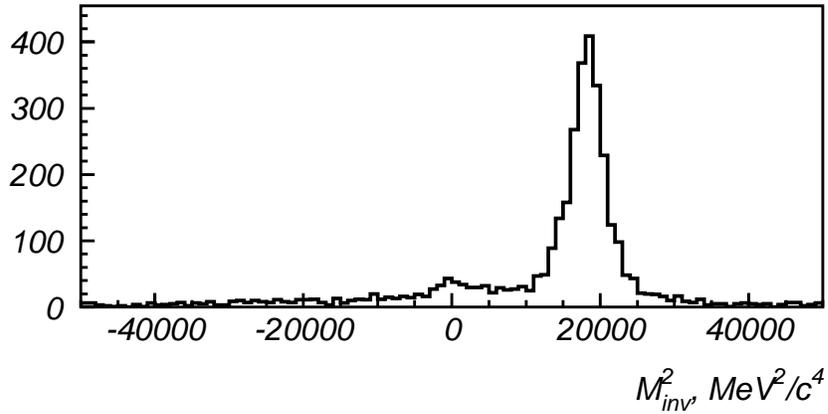}
\vspace{-6cm}
\caption{{\bf $M^2_{inv}$} distribution for selected experimental data.}
\label{fig:expn}
\end{center}
\end{figure}

\hspace*{\parindent}The distribution over the parameter $M^2_{inv}$ 
for selected events is presented in Fig.~\ref{fig:expn}. Events 
from the decay $\phi \to \eta \gamma$ produce a peak near the 
$\pi^0$ mass squared. Other events surviving after all the cuts are coming
from the following processes:
\begin{itemize}
\item{\boldmath $\phi \to \eta\gamma, \eta \to \pi^+\pi^-\gamma$}. 
For such events the value of $M_{inv}$ should be obviously zero and a
corresponding peak can be seen in Fig. \ref{fig:expn}.
\item{\boldmath $e^+e^- \to \omega\pi^0 \to \pi^+\pi^-\pi^0\pi^0$}
\item{\boldmath $\phi \to \pi^+\pi^-\pi^0$}. Most of the events from
this decay are suppressed by the cut on the total energy of the charged
particles $\varepsilon_{\pi^+\pi^-}<520~MeV$, but some small fraction
could survive.
\item{\boldmath $\phi \to K_SK_L$}
\item{beam background, the photon conversion at the vacuum tube}
\end{itemize}

\begin{figure}
\begin{center}
\includegraphics[width=0.9\textwidth]{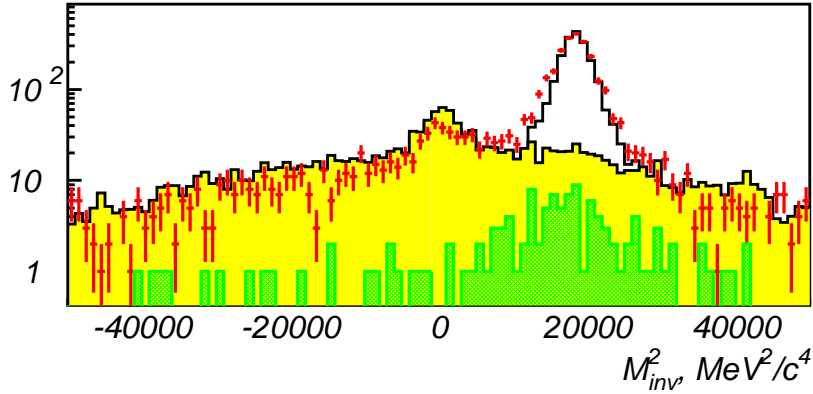}
\vspace{-6.5cm}
\caption{{\bf $M^2_{inv}$} distribution for data (points with errors)
together with the sum of simulated background processes (hatched histogram). 
The contribution of $\phi\to\pi^+\pi^-\pi^0$ is shown by the solid hatched 
histogram.}
\label{fig:expb}
\end{center}
\end{figure}

Simulation of the first three background processes used 
the GEANT based CMD-2 simulation code \cite{mcpre}. It is
difficult to study two last sources of background using simulation,
therefore their behaviour was studied with the help of experimental 
data using the selection $R_{min}~>~0.3~cm$ instead of $R_{min}~<~0.2~cm$.
Figure~\ref{fig:expb}  presents the distribution over 
$M^2_{inv}$ for simulated events of background processes after
applying selection criteria. The fraction of each background process in
the total number of simulated events corresponds to its
cross section, the fraction of background with large distances from the beam
has been also normalized. The sum of background distributions 
is shown by the hatched histogram in Fig.~\ref{fig:expb}. The logarithmic scale
was chosen for better demonstration of the background.  
The background simulation is in reasonable agreement with the data 
supporting our description of the background processes. 
sources is correct. The background from the $\phi \to \pi^+\pi^-\pi^0$ decay 
(see the solid hatched histogram in Fig.~\ref{fig:expb}) is rather small but 
unpleasant since its distribution for events surviving the selection has a
maximum near that of the signal. 

The number of $\phi \to \eta \gamma$ events in the mode $\eta \to
\pi^+\pi^-\pi^0$ was determined by the following procedure:
\begin{itemize}
\item The distribution over $M_{inv}^2$ for each background process was fit by
a smooth function to fix its shape. 
\item The distribution over $M_{inv}^2$ for all the data (Fig.~\ref{fig:expn}) 
was fit by the sum of background functions excluding the signal region.
\item After that the function describing the shape of the signal was
determined by the fit including the signal region and taking the sum of
background functions with fixed parameters.
\item As a result of previous steps one had the functions describing
the shape of signal and each background component.
For each beam energy the number of 
$\phi \to \eta\gamma \to \pi^+\pi^-\pi^0 \gamma$ events was determined by
fitting the distribution over $M_{inv}^2$ by the sum 
of these functions with the fixed shape and the number of events for signal 
and background processes as free parameters.  
For the $\phi \to \pi^+\pi^-\pi^0$ background 
the detection efficiency was determined from the simulation
and the number of events
at each energy  was calculated according to 
the energy dependence of the cross section of the reaction
$e^+e^- \to \phi \to \pi^+\pi^-\pi^0$ \cite{kuzpi3}. The total
background of the decay mode $\phi \to \pi^+\pi^-\pi^0$ is less
than 2\%.
\end{itemize}

The resulting number of $\eta\gamma$ events and the obtained
cross section are presented in Table \ref{tab:rez}. 

At each energy the cross section was calculated as
$$\sigma = \frac{N_{\eta\gamma}}{\mathcal{L}\cdot (1+\delta)
\cdot \varepsilon \cdot B_{\eta\to\pi^+\pi^-\pi^0}},$$
where $\mathcal{L}$ is the integrated luminosity at this energy,
$\delta$ is a radiative correction for the process $\phi \to \eta
\gamma$, $\varepsilon$ is the detection efficiency, 
$B_{\eta\to\pi^+\pi^-\pi^0}$ = $0.231 \pm 0.005$ \cite{pdg}.

\begin{figure}[htb]
\includegraphics[width=0.9\linewidth]{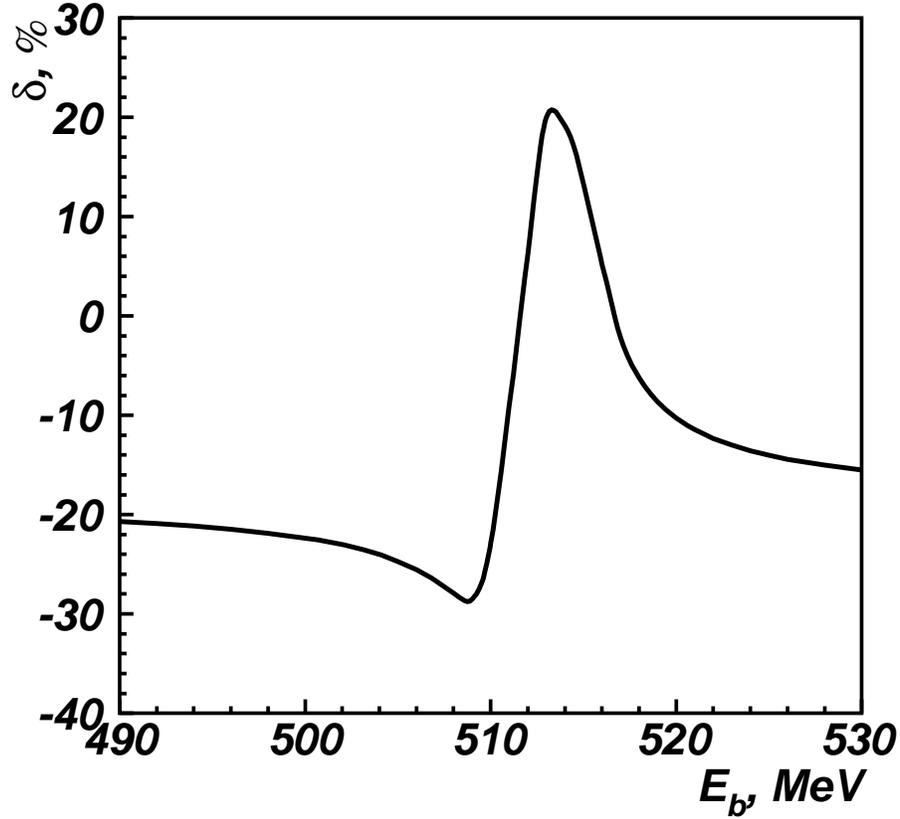}
\caption{Radiative correction $\delta$ for $\phi \to \eta \gamma$
vs energy}
\label{fig:radcor}
\end{figure}

The radiative correction was calculated according to \cite{kurfad} and
its behaviour in the $\phi$-meson energy range is presented in 
Fig.~\ref{fig:radcor}.

The detection efficiency was determined using simulation of
$\phi \to \eta \gamma, \eta \to \pi^+\pi^-\pi^0$ and the value 
$\varepsilon = 0.247 \pm 0.001 \pm 0.007$ was obtained. 
The systematic uncertainty of the efficiency was estimated
by comparing results of the simulation of nuclear interactions by different
methods (FLUKA and GHEISHA) \cite{mcpre}. For simulation of the decay
$\eta\to\pi^+\pi^-\pi^0$ instead of standard GEANT tools which assume a 
constant matrix element we used the following matrix element parametrisation 
from \cite{mont} 
$$|M(x,y)|^2 \propto (1 + ay + by^2 + cx + dx^2 + exy).$$
Here $x = \sqrt{3}(T_+ - T_-)/Q$, $y = (3T_0/Q) - 1$,\\
where $T_+, \ T_-$ and $T_0$ are the kinetic energies of the 
$\pi^+, \ \pi^-$ and $\pi^0$ in the $\eta$ rest frame, and
$Q = T_+ + T_- + T_0$. The parameter values were taken from recent 
Crystal Barrel data \cite{abe}.

The cross section of the process $e^+e^- \to  \phi \to \eta \gamma$ was 
parametrized by a sum of the 
$\phi$-meson amplitude and an interfering term with the tails of the $\rho$
and $\omega$ resonances:

$\sigma_{\eta\gamma} = \frac{F_{\eta\gamma}(s)}{s^{3/2}}\cdot
|A_\phi + A|^2,$

$A_\phi = \frac{m_\phi^{5/2}\Gamma_\phi\sqrt{\sigma^0_\phi/F_{\eta\gamma}
(m_\phi^2)}}{s-m_\phi^2+i\sqrt{s}\Gamma_\phi(s)},$

\noindent
where $s=4E_b^2$, $m_\phi, \ \Gamma_\phi, \ \sigma^0_\phi$ are the mass, 
width and peak cross section ($s=m_\phi^2$) for the $\phi$-meson, 
and $F_{\eta\gamma}(s)
= (\sqrt{s}(1 -m_\eta^2/s))^3$ is a function describing the
dynamics of the $\phi \to \eta \gamma$ decay including the phase space.

\begin{table}
\caption{The c.m.energy, integrated luminosity, number of $\eta\gamma$ events,
radiative correction  and cross section of $\phi \to \eta\gamma$}
\vspace{.3cm}
\begin{tabular}{cccccc}
\hline
{\bf No.}&{\boldmath $\sqrt{s}, \ MeV$}&{\boldmath $\int\mathcal{L}dt,
\ nb^{-1}$}&{\boldmath $N_{\eta\gamma}$}&{\boldmath $\delta$}&
{\boldmath $\sigma_{\eta\gamma}, \ nb$}\\
\hline
 1&$ 985.80\pm0.20$&$ 93.4\pm 0.5$&$  3.2\pm 3.8$&-0.21&$ 0.8 \pm 0.9$\\
 2&$1004.80\pm0.20$&$ 98.8\pm 0.5$&$  7.6\pm 4.3$&-0.24&$ 1.8 \pm 1.0$\\
 3&$1011.44\pm0.20$&$108.9\pm 0.6$&$ 11.8\pm 4.7$&-0.25&$ 2.5 \pm 1.0$\\
 4&$1016.14\pm0.20$&$123.4\pm 0.6$&$110.0\pm11.4$&-0.28&$21.7 \pm 2.3$\\
 5&$1017.08\pm0.06$&$177.8\pm 0.7$&$192.1\pm14.9$&-0.28&$26.3 \pm 2.0$\\
 6&$1018.04\pm0.06$&$226.2\pm 0.8$&$347.4\pm20.0$&-0.28&$37.4 \pm 2.2$\\
 7&$1018.88\pm0.04$&$ 84.6\pm 0.5$&$170.3\pm14.1$&-0.27&$48.4 \pm 4.0$\\
 8&$1019.04\pm0.04$&$ 63.8\pm 0..4$&$128.9\pm12.3$&-0.27&$48.5 \pm 4.6$\\
 9&$1019.20\pm0.04$&$103.4\pm 0.5$&$239.9\pm16.6$&-0.26&$55.0 \pm 3.8$\\
10&$1019.84\pm0.04$&$ 87.0\pm 0.5$&$185.4\pm14.5$&-0.24&$49.2 \pm 3.8$\\
11&$1020.08\pm0.04$&$170.7\pm 0.7$&$340.5\pm19.6$&-0.22&$44.9 \pm 2.6$\\
12&$1020.72\pm0.04$&$ 40.8\pm 0.3$&$ 57.6\pm 8.5$&-0.19&$30.6 \pm 4.5$\\
13&$1020.98\pm0.06$&$143.8\pm 0.6$&$207.7\pm15.7$&-0.17&$30.5 \pm 2.3$\\
14&$1021.80\pm0.06$&$119.5\pm 0.6$&$102.1\pm11.4$&-0.11&$16.8 \pm 1.9$\\
15&$1022.72\pm0.06$&$ 77.6\pm 0.5$&$ 56.4\pm 8.3$&-0.04&$13.3 \pm 2.0$\\
16&$1028.26\pm0.10$&$ 84.7\pm 0.5$&$  9.2\pm 4.0$& 0.28&$ 1.5 \pm 0.6$\\
17&$1033.70\pm0.18$&$ 55.7\pm 0.5$&$  3.2\pm 3.5$&-0.01&$ 1.0 \pm 1.1$\\
18&$1038.96\pm0.12$&$ 22.9\pm 0.5$&$  2.5\pm 3.5$&-0.09&$ 2.1 \pm 2.9$\\
\hline
\end{tabular}
\label{tab:rez}
\end{table}

\vspace{5mm}
The energy dependence of the $\phi$ meson width was written as\\
$$\Gamma_\phi(s) = \Gamma_\phi\cdot\Biggl(B_{K^+K^-}\frac{m_\phi^2
F_{K^+K^-}(s)}{s\cdot F_{K^+K^-}(m_\phi^2)} +
B_{K_SK_L}\frac{m_\phi^2F_{K_SK_L}(s)}{s\cdot F_{K_SK_L}(m_\phi^2)} +$$ 
$$B_{\pi^+\pi^-\pi^0}\frac{\sqrt{s}F_{\pi^+\pi^-\pi^0}(s)}{m_\phi
F_{\pi^+\pi^-\pi^0}(m_\phi^2)} + B_{\eta\gamma}\frac{F_{\eta\gamma}(s)}
{F_{\eta\gamma}(m_\phi^2)} \Biggr),$$
where $B_{K^+K^-}$, $B_{K_SK_L}$, $B_{\pi^+\pi^-\pi^0}$ and
$B_{\eta\gamma}$ are branching ratios of the $\phi$ meson decay into 
corresponding modes; $F_{K\bar{K}}(s) = (s/4 - m_K^2)^{3/2}$, and for the 
$F_{\pi^+\pi^-\pi^0}(s)$ calculation the model assuming the decay
$\phi(\omega) \to \rho\pi \to \pi^+\pi^-\pi^0$ was used.

\begin{figure}[htb]
\includegraphics[width=0.9\linewidth]{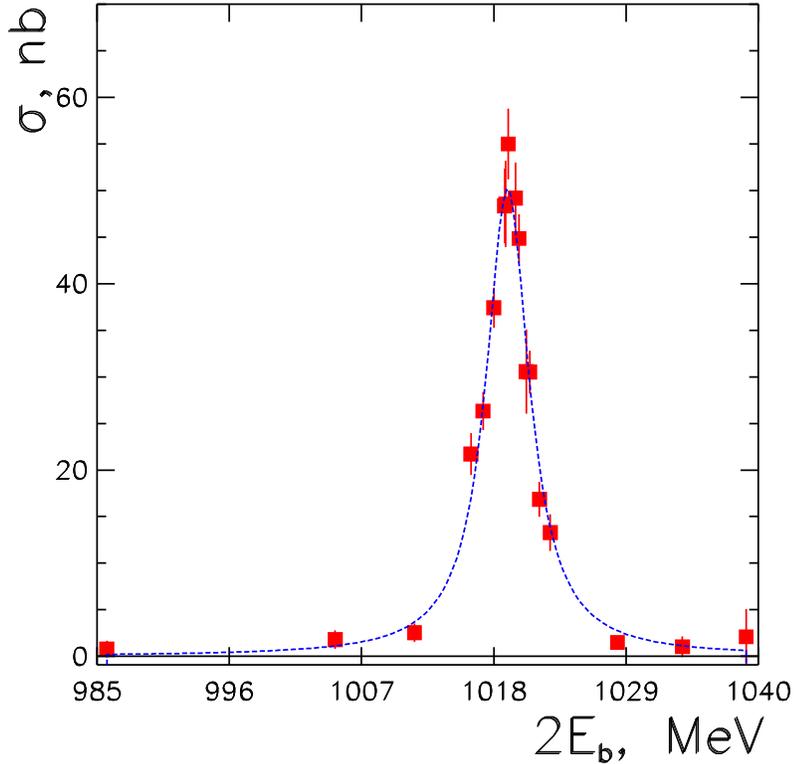}
\vspace{-3cm}
\caption{The cross section of $e^+e^- \rightarrow \phi \rightarrow
\eta\gamma$ with the fit function.}
\label{fig:sigphi}
\end{figure}

Fitting the energy dependence of the cross section of $\phi \to
\eta\gamma$ (Fig.~\ref{fig:sigphi}) the $\phi$ meson mass and peak 
cross section were determined, while $\Gamma_\phi$ was 
fixed at the world average value \cite{pdg}. 
The following results were obtained when the interfering term $A=0$:\\
$\sigma_\phi^0=49.8\pm1.2\pm2.4$ nb;
$m_\phi=1019.38\pm0.07\pm0.08$ MeV;
$\chi^2/d.f.=18.4/16$.

If $A$ is calculated accurately taking into account the $\rho$ and 
$\omega$ meson contributions, the value of
$\sigma_\phi^0$ is changed by only 1\%.

The systematic uncertainty of the peak cross section is estimated to be 4.8\% 
and includes the following contributions:
\begin{itemize}
\item{luminosity determination (2\%);}
\item{the uncertainty of the value of $B(\eta\to\pi^+\pi^-\pi^0)$ (2.2\%);}
\item{determination of the detection efficiency (2.8\%);}
\item{uncertainty of the background subtraction (2.5\%)}
\end{itemize}

The systematic uncertainty of the mass was estimated to be 80 keV from the 
difference in the c.m.energy obtained from two different methods of the beam 
energy determination (one used the momenta of charged kaons and the other was
based on the analysis of the magnetic field of the collider \cite {prep99}).   

Taking the value $\Gamma_{ee}/\Gamma_{tot} = (2.99 \pm 0.08) \times
10^{-4}$ from \cite{pdg}, one obtains the following value of the branching 
ratio corresponding to the peak cross section above:
$$B_{\phi\to\eta\gamma} = (1.18 \pm 0.03 \pm 0.06)\%,$$
where the first error is statistical and the second one is systematic.
The latter includes contributions from the uncertainties 
of the peak cross section (4.8\%) and that of 
$\Gamma_{ee}/\Gamma_{tot}$ (2.7\%).

\section{Conclusions}

\hspace*{\parindent}
Using a data sample of about 3.5 million $\phi$ meson decays
the CMD-2 collaboration studied the decay 
$\phi \to \eta\gamma, \eta \to \pi^+\pi^-\pi^0$. 
About 2200 events of this decay have been selected.
The $\phi$ meson mass 
was determined to be $(1019.38 \pm 0.07 \pm 0.08) MeV$ and agrees
with other measurements \cite {pdg}. The branching ratio of the decay
was measured with high precision to be $(1.18 \pm 0.03 \pm 0.06)\%$ 
and is consistent with the world average value $(1.26 \pm 0.06)\%$ \cite{pdg} 
as well as with another accurate determination of this quantity recently
reported by the SND collaboration $(1.246 \pm 0.025 \pm 0.057)\%$ \cite{snd}.
It is the first measurement of the $\phi \to \eta\gamma$ decay rate 
using the charged mode of the $\eta$ meson decay. Analysis of a five
times bigger data sample is in progress.

\section{Acknowledgements}

\hspace*{\parindent}
The authors are grateful to M.Benayoun, V.N.Ivanchenko and
A.A.Salnikov for useful discussions.

\end{document}